\documentstyle[preprint,aps]{revtex}
\tightenlines  
\input{psfig}
\newcommand {\ebar}{\hbox{{\em E}\kern-0.5em\lower-0.1ex\hbox{/}}} 
\def\simlt{\stackrel{<}{{}_\sim}}
\def\simgt{\stackrel{>}{{}_\sim}}
\begin{document}
\draft

\title{Search for Narrow Diphoton Resonances  and for
$\gamma\gamma+W/Z$ Signatures in $p\bar{p}$ Collisions at $\sqrt{s}=1.8$~TeV}

\date{\today}

\maketitle


\font\eightit=cmti8
\def\r#1{\ignorespaces $^{#1}$}
\hfilneg
\begin{sloppypar}
\noindent
T.~Affolder,\r {23} H.~Akimoto,\r {45}
A.~Akopian,\r {37} M.~G.~Albrow,\r {11} P.~Amaral,\r 8  
D.~Amidei,\r {25} K.~Anikeev,\r {24} J.~Antos,\r 1 
G.~Apollinari,\r {11} T.~Arisawa,\r {45}
% ...
\end{sloppypar}
\vskip .026in
\begin{center}
(CDF Collaboration)
\end{center}

\vskip .026in
\begin{center}
\r 1  {\eightit Institute of Physics, Academia Sinica, Taipei, Taiwan 11529, 
Republic of China} \\
\r 2  {\eightit Argonne National Laboratory, Argonne, Illinois 60439} \\
\r 3  {\eightit Istituto Nazionale di Fisica Nucleare, University of Bologna,
I-40127 Bologna, Italy} \\
\r 4  {\eightit Brandeis University, Waltham, Massachusetts 02254} \\
\r 5  {\eightit University of California at Davis, Davis, California  95616} \\
\r 6  {\eightit University of California at Los Angeles, Los 
Angeles, California  90024} \\  
\r 7  {\eightit Instituto de Fisica de Cantabria, CSIC-University of 
Cantabria,
39005 Santander, Spain} \\
\r 8  {\eightit Enrico Fermi Institute, University of Chicago, Chicago, 
Illinois 60637} \\
\r 9  {\eightit Joint Institute for Nuclear Research, RU-141980 Dubna, Russia}
\\
\r {10} {\eightit Duke University, Durham, North Carolina  27708} \\
\r {11} {\eightit Fermi National Accelerator Laboratory, Batavia, Illinois 
60510} \\
\r {12} {\eightit University of Florida, Gainesville, Florida  32611} \\
\r {13} {\eightit Laboratori Nazionali di Frascati, Istituto Nazionale di 
Fisica
               Nucleare, I-00044 Frascati, Italy} \\
\r {14} {\eightit University of Geneva, CH-1211 Geneva 4, Switzerland} \\
\r {15} {\eightit Glasgow University, Glasgow G12 8QQ, United Kingdom}\\
\r {16} {\eightit Harvard University, Cambridge, Massachusetts 02138} \\
\r {17} {\eightit Hiroshima University, Higashi-Hiroshima 724, Japan} \\
\r {18} {\eightit University of Illinois, Urbana, Illinois 61801} \\
\r {19} {\eightit The Johns Hopkins University, Baltimore, Maryland 21218} \\
\r {20} {\eightit Institut f\"{u}r Experimentelle Kernphysik, 
Universit\"{a}t Karlsruhe, 76128 Karlsruhe, Germany} \\
\r {21} {\eightit Center for High Energy Physics: Kyungpook National
University, Taegu 702-701; Seoul National University, Seoul 151-742; and
SungKyunKwan University, Suwon 440-746; Korea} \\
\r {22} {\eightit High Energy Accelerator Research Organization (KEK), 
Tsukuba,
Ibaraki 305, Japan} \\
\r {23} {\eightit Ernest Orlando Lawrence Berkeley National Laboratory, 
Berkeley, California 94720} \\
\r {24} {\eightit Massachusetts Institute of Technology, Cambridge,
Massachusetts  02139} \\   
\r {25} {\eightit University of Michigan, Ann Arbor, Michigan 48109} \\
\r {26} {\eightit Michigan State University, East Lansing, Michigan  48824} \\
\r {27} {\eightit University of New Mexico, Albuquerque, New Mexico 87131} \\
\r {28} {\eightit The Ohio State University, Columbus, Ohio  43210} \\
\r {29} {\eightit Osaka City University, Osaka 588, Japan} \\
\r {30} {\eightit University of Oxford, Oxford OX1 3RH, United Kingdom} \\
\r {31} {\eightit Universita di Padova, Istituto Nazionale di Fisica 
          Nucleare, Sezione di Padova, I-35131 Padova, Italy} \\
\r {32} {\eightit University of Pennsylvania, Philadelphia, 
        Pennsylvania 19104} \\   
\r {33} {\eightit Istituto Nazionale di Fisica Nucleare, University and Scuola
               Normale Superiore of Pisa, I-56100 Pisa, Italy} \\
\r {34} {\eightit University of Pittsburgh, Pittsburgh, Pennsylvania 15260} \\
\r {35} {\eightit Purdue University, West Lafayette, Indiana 47907} \\
\r {36} {\eightit University of Rochester, Rochester, New York 14627} \\
\r {37} {\eightit Rockefeller University, New York, New York 10021} \\
\r {38} {\eightit Rutgers University, Piscataway, New Jersey 08855} \\
\r {39} {\eightit Texas A\&M University, College Station, Texas 77843} \\
\r {40} {\eightit Texas Tech University, Lubbock, Texas 79409} \\
\r {41} {\eightit Institute of Particle Physics, University of Toronto, Toronto
M5S 1A7, Canada} \\
\r {42} {\eightit Istituto Nazionale di Fisica Nucleare, University of Trieste/
Udine, Italy} \\
\r {43} {\eightit University of Tsukuba, Tsukuba, Ibaraki 305, Japan} \\
\r {44} {\eightit Tufts University, Medford, Massachusetts 02155} \\
\r {45} {\eightit Waseda University, Tokyo 169, Japan} \\
\r {46} {\eightit University of Wisconsin, Madison, Wisconsin 53706} \\
\r {47} {\eightit Yale University, New Haven, Connecticut 06520} \\
\r {(\ast)} {\eightit Now at Carnegie Mellon University, Pittsburgh,
Pennsylvania  15213} \\
\r {(\ast\ast)} {\eightit Now at Northwestern University, Evanston, Illinois 
60208}
\end{center}

\newpage

\begin{abstract}
We present results of searches for diphoton resonances produced
both inclusively and also in association with a vector boson ($W$ or $Z$) 
using $\rm
100~pb^{-1}$\ of $p \bar p$\ collisions using the CDF detector.  We
set upper limits on the product of cross section times branching ratio
for both $p \bar p \rightarrow \gamma\gamma + X$ and $p \bar p \rightarrow
\gamma\gamma + W/Z$. 
Comparing the inclusive production to the expectations from heavy
sgoldstinos we derive limits on the supersymmetry-breaking scale $\sqrt{F}$
in the TeV range, depending on the sgoldstino mass and the choice of other
parameters.
Also,
 using a NLO prediction for the associated production
of a Higgs boson with a $W$\ or $Z$ boson, we set an
upper limit on the branching ratio for $H \rightarrow \gamma\gamma$.
Finally, we set a lower limit on the mass of a `bosophilic' Higgs boson
(e.g. one which couples only to $\gamma$, $W$, and $Z$\ bosons with
standard model couplings) of $\rm 82~GeV/c^2$\ at 95\% confidence level.
\end{abstract}


%
%

\section{Introduction}

Many processes in extensions of the standard model (SM) result in
final-state signatures involving two vector gauge bosons, $VV+X$,
where $V$ is either a $W$, $Z$, or photon. The signature of high mass
photon pairs is attractive for searches for new physics as
the photon is the lightest gauge boson, and hence might be more easily
produced in decays of new particles. In addition, the photon, being stable,
does not decay into many different final states as do the $W$ and $Z$.
The dominant SM background  process, the production of very massive photon
pairs ($M_{\gamma\gamma}\simgt 100$ GeV/c$^2$), is small 
compared to the cross-sections for
producing new strongly-interacting 
states via quark-antiquark annihilation, making this an
attractive channel in which to search for new particles or
interactions. Examples of possible sources of high mass diphoton pairs
include sgoldstino production~\cite{zwirh}, interaction terms arising from
extra spatial dimensions~\cite{hall}, a new interaction at a high scale
manifesting itself as a $q\bar q\to \gamma\gamma$\ contact
interaction~\cite{contact}, a `bosophilic' Higgs
boson~\cite{haber79,stange94,diaz94,akeroyd96}, or a heavy analog 
of the $\pi^0$\ that
also does not couple to fermions~\cite{technipizero}.  In this paper we
focus on the production of sgoldstinos and Higgs bosons and their decay
into two photons.

\par
Models with spontaneous breaking of global supersymmetry require 
  a massless and neutral spin-$\frac{1}{2}$
 particle, the goldstino (${\tilde G}$).
 When gravitation is added and supersymmetry
 is realized locally the gauge particle, the graviton, has a
 spin-$\frac{3}{2}$ partner, the gravitino, which acquires a mass
 while the goldstino is absorbed~\cite{grav}. Goldstinos (R-odd) have
 supersymmetric partners called sgoldstinos (R-even) which are
 expected to be a part of the effective theory at the weak scale if
 gravitinos are very light ($\simlt 10^{-3}$ eV/c$^2$). The simplest
 model considers two neutral spin-0 states: S (CP-even) and P (CP-odd),
 for which we use the generic symbol $\phi$. The mass for these states
 is completely arbitrary and although initially signals were studied in
 the limit of vanishing masses~\cite{vanish}, we follow the suggestions
 of Ref.~\cite{zwirh} and concentrate on massive 
 sgoldstinos, $M_\phi ={\cal O}(100~ \rm{GeV/c^2})$. 
 The production of sgoldstinos is dominated by
 the gluon-gluon fusion process~\cite{zwirh} while their decay is
 dominated by two-body decays into a pair of gluons, goldstinos,
 photons, $W$'s, $Z$'s and top quarks. The corresponding branching
 ratios have been calculated~\cite{zwire} for two specific choices of
 parameters, the branching ratio into two photons being of the order
 of a few percent. Limits on the supersymmetry-breaking scale $\sqrt{F}$
 have been set by the DELPHI Collaboration~\cite{delphi} for sgoldstino
 masses up to about 200 GeV/c$^2$.  We take advantage here of the
 higher energy reached at the Tevatron to extend the search to much larger
 masses.
\par

\begin{figure}
\centerline{
\psfig{figure=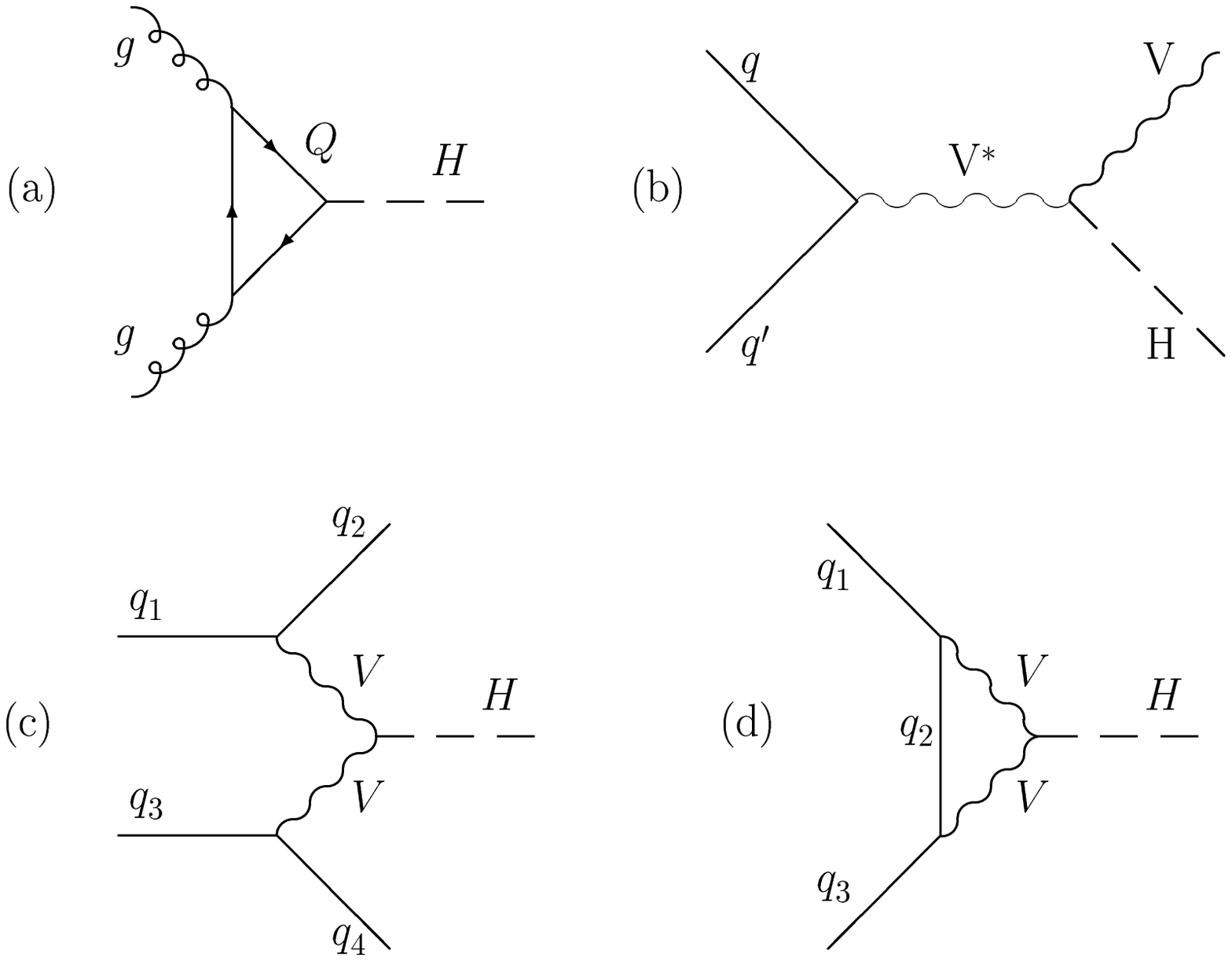,height=16cm,width=16cm}}
\caption{ Diagrams for production of a Higgs Boson in $p \bar p$\ collisions: 
(a) gluon-gluon fusion, (b) associated production with a vector boson,
(c) and (d) vector boson fusion. In the bosophilic models the
 gluon-gluon fusion diagram is suppressed. } \label{fig:higgs_prod}
\end{figure}

\begin{figure}
\centerline{
\psfig{figure=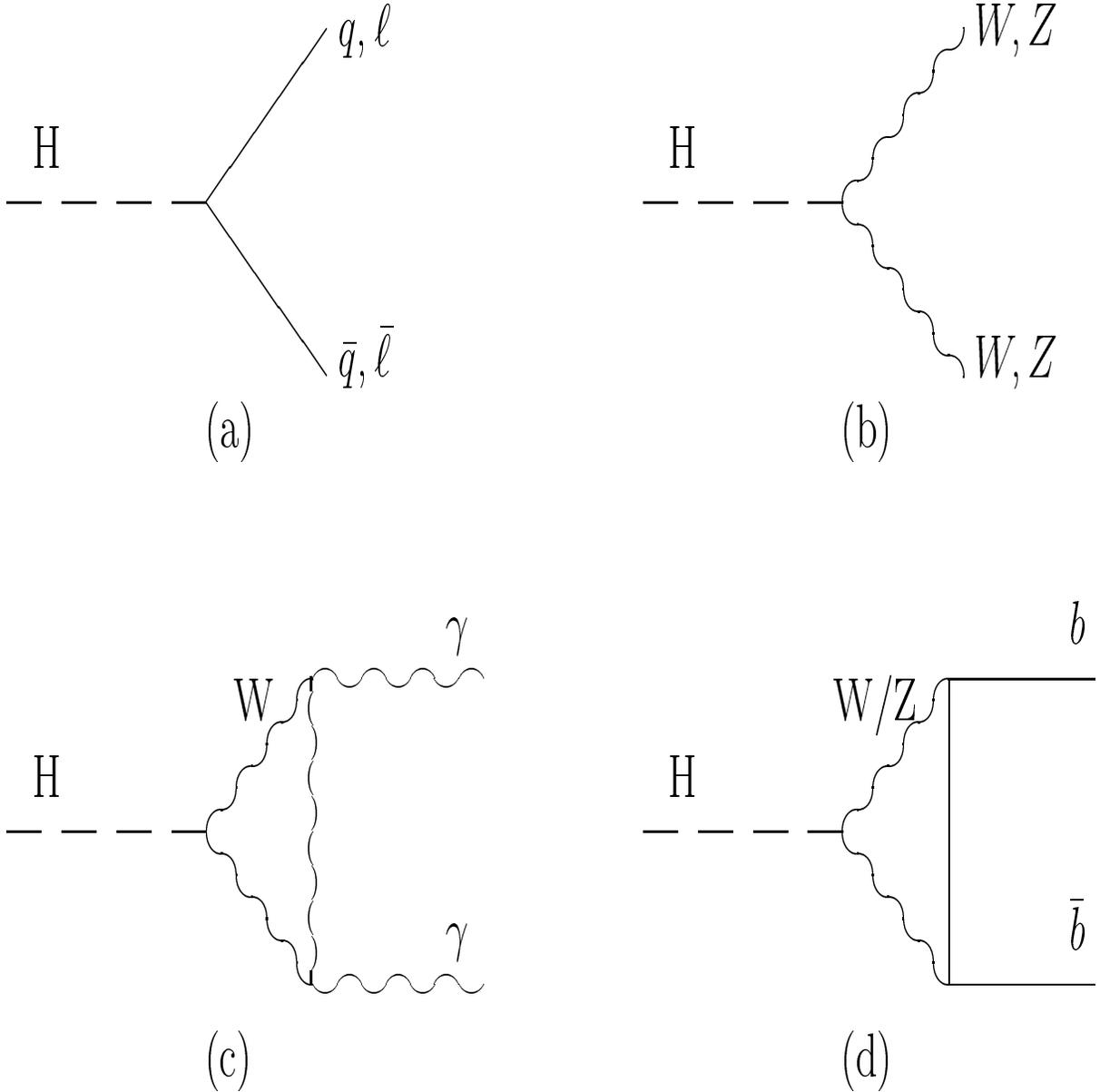,height=16cm,width=16cm}}
\caption{ Diagrams for the decay of a Higgs Boson to: (a) a quark or lepton
pair, (b) vector boson 
pairs (WW/ZZ),  (c) via a loop to $\gamma\gamma$, and (d) via a loop to
$b \bar b$.  For a bosophilic Higgs, the decay to $b \bar b$\ is
suppressed relative to $\gamma\gamma$. }
\label{fig:higgs_decay}
\end{figure}

There are also models in which a Higgs boson could decay into two
photons with a branching ratio much larger than predicted in the
standard model.  Figure~\ref{fig:higgs_prod} shows the dominant
diagrams for production of a standard model Higgs boson ($H$) in $p \bar p$\
collisions.  The total production cross section is dominated by the
gluon-gluon fusion process, and has a value of approximately 1~pb for
$M_H \sim 100$ GeV/c$^2$~\cite{stange94,higgs_nlo_xsec}.
Figure~\ref{fig:higgs_decay} shows the dominant decay diagrams for a
SM Higgs with mass less than $\sim$130~GeV/c$^2$. The dominant decay mode
of the $H$\ in this mass range is
\mbox{$H\to b\bar b$}, with the branching ratio to $\gamma\gamma$
being on the order of $10^{-4}$.  At higher masses, the decays to
vector boson pairs $WW$ and $ZZ$ dominate.  However, some models
beyond the standard model introduce anomalous
couplings~\cite{anomalous_higgs} or additional Higgs multiplets
~\cite{stange94,akeroyd96}, enhancing the coupling to photons or
suppressing the coupling to fermions.  The result is a low-mass Higgs
boson with significantly increased branching ratio to two photons.  In
the bosophilic models, the coupling to fermions at tree level is set
to zero while maintaining the SM coupling to vector bosons.  Although
the decay to two photons proceeds through a higher-order loop diagram,
it is the dominant decay for $M_H<M_W$.  For $M_H>M_W$\ the decay $H
\to WW^*$\ becomes dominant.  Since the bosophilic Higgs has no 
coupling to fermions, 
the gluon-gluon fusion production mechanism is lost and the
dominant production mode in $p\bar p$\ collisions at
$\sqrt{s}=1.8$~TeV is associated production with a $W$\ or $Z$\ boson.
For $M_H = 80$~ GeV/c$^2$, the total associated production cross
section is about 0.8~pb.  The limit set in this paper uses the
branching ratios of reference~\cite{stange94}.

Limits on the mass and branching ratios of a bosophilic Higgs boson
have been set by the OPAL Collaboration assuming SM production of $Z
H$\, with a lower limit on $M_H$\ of 96.2~ GeV/c$^2$\ at 95\%
confidence level
(C.L.)~\cite{opal_ggx}. More recently, a limit of 100.7~ GeV/c$^2$\
at 95\% C.L.~\cite{aleph_ggx} has been set by the ALEPH Collaboration.
The D0 Collaboration at Fermilab has set a lower limit of 78.5~
GeV/c$^2$\ at 95\% C.L.~\cite{D0_ggjj} in a search at the Tevatron for
$W H$\ and $Z H$\ production.

In this paper we describe a search for departures from SM expectations
for both inclusive high-mass $\gamma\gamma$\ production and also
$\gamma\gamma$\ production in association with a $W$\ or $Z$\ boson.
This search uses $100\pm 4$~ pb$^{-1}$\ of data collected between 1992
and 1995 with the Collider Detector at Fermilab (CDF).  The photon
selection criteria for this analysis were optimized to remain
efficient for very high energy photons. The analysis is complementary to
the previous QCD diphoton cross section measurement~\cite{diphoxsec}.  
In this present analysis, the photon selection criteria have been
optimized for high efficiency, taking advantage of the smaller jet fake
background rate at high $E_T$.
 The analysis is also complementary to the recent diphoton + X search
analysis~\cite{ggmet} which searched for non-resonant diphoton
signatures, such as
$ee\gamma\gamma\ebar_T$, that might arise in gauge-mediated
supersymmetric models.

\section{ The CDF Detector}

We briefly describe the CDF detector, which is described in 
detail elsewhere~\cite{cdf}.  The magnetic spectrometer consists of
three tracking devices immersed in the 1.4~T field of a 3~m-diameter
5~m-long super-conducting solenoid.  The magnetic field and three
tracking devices are all arranged with their principal axis parallel
to the proton beam direction ($z$-axis)~\cite{coordinates}.  The
tracking device closest to the beam line is a four-layer silicon
micro-strip vertex detector (SVX), used to find secondary vertices,
with layers at radii from 2.8~cm to 7.9~cm~\cite{SVX}.  Surrounding the
SVX is a set of time projection chambers (VTX) which identifies the
$p\bar p$ interaction point(s) along the beam axis 
with a series of $r-z$\ measurements out to a 
radius of 22~cm.  The central tracking chamber (CTC) is a 3.5~m-long
84 layer drift chamber surrounding the VTX.    The CTC wires, ranging in
radius from 31.0~cm to 132.5~cm, are arranged in 5 superlayers of
axial wires alternating with 4 superlayers of stereo wires.  The
calorimeter, which is constructed in projective electromagnetic and
hadronic towers, consists of the central barrel ($|\eta|<1.1$) which
surrounds the solenoid, the end-plugs ($1.1<|\eta|<2.4$) which form
the magnet poles and the forward calorimeters ($2.4<|\eta|<4.2$).
 Wire chambers
with cathode strip readout (CES) are located at shower maximum in the
central electromagnetic calorimeter.  These chambers provide a two-dimensional 
shower profile which is used to discriminate on a
statistical basis between photons and $\pi^0$\ backgrounds.
Additional statistical discrimination is provided by exploiting the 
difference in
conversion probability for single photons and pairs from $\pi^0$\ decays
in the 1 radiation length
of the coil.  The presence of a conversion is detected using wire
chambers (CPR) located between the coil and the central calorimeter.
The central muon chambers ($|\eta|<1.1$) are located outside the central
calorimeter to detect particles penetrating the calorimeter.

\section{ Diphoton Event Selection }

Photons are identified as a narrow shower in the electromagnetic
calorimeter with no associated high-$P_T$ charged particle track.  The
energy in the hadronic calorimeter and adjoining regions of the
electromagnetic calorimeter must be small to reject jet backgrounds.
For high-$E_T$\ photons there is a background from $\pi^0 \to
\gamma\gamma$\ decays where both photons are very close together.
  
The candidate $\gamma\gamma$\ events must pass the diphoton
requirements of the three-level CDF trigger.  The first hardware level
requires two central electromagnetic calorimeter trigger towers with
$E_T>4$~GeV.  The second hardware level requires two central
electromagnetic trigger clusters~\cite{trig_cluster} with $E_T>16$~GeV
and a ratio of hadronic to electromagnetic energy satisfying
$E_T(HAD)/E_T(EM)<0.125$.  In the third trigger level, electromagnetic
clusters~\cite{cluster} are found using the offline reconstruction
algorithm and the 16~GeV threshold is re-applied to the recalculated
transverse energy of the new cluster.
\par
Offline event selection requires at least two central electromagnetic
clusters each satisfying the following requirements: $E_T>22$~GeV, no
track pointing at the cluster (or one track with $P_T<1$~GeV/c), pulse
height and cluster shape in the central electromagnetic strip chamber
(CES) consistent with a photon (to reject $\pi^0$'s and cosmic rays),
no additional CES cluster in the same $15^{\circ}$\ azimuthal section
of the calorimeter (to reject $\pi^0$'s), and minimal energy deposited
in the hadronic calorimeter towers behind the cluster.
 
Isolation requirements, based on track and calorimeter activity in an
$\eta-\phi$\ cone with radius $\Delta R \equiv
\sqrt{(\Delta\phi)^2+(\Delta\eta)^2} = 0.4$\ around the cluster, are
used to reduce backgrounds from jets: $\Sigma P_T(Tracks) <5.0$~GeV/c
and $(E_T(\Delta R<0.4)-E_T(Cluster))<2.0$~GeV. The calorimeter
isolation energy is corrected for leakage from the cluster and for
pile-up from multiple interactions.  The efficiency of the calorimeter
isolation requirement is studied as a function of $E_T$\ using a
sample of electrons from $W \rightarrow e\nu$\ events.  The efficiency
for electrons with $30<E_T<100$~GeV is $94.0\pm0.1\%$\ and for
electrons with $100<E_T<200$~GeV is $94.9\pm0.6\%$.  Two requirements
reject backgrounds from cosmic rays: there must be at least one
reconstructed primary vertex within $\pm60$~cm of the center of the
interaction region along the beam direction, and all energy measured
in the central hadronic calorimeters is required to be in time with
the collision.

The efficiency to identify a photon passing the above isolation
criteria within the fiducial region of the central calorimeter is
measured using a control sample of electrons from $Z^0$\ decay to be
$84\pm4\%$.  The combined diphoton and event selection efficiency is
$63\pm6\%$\ (the geometric factor due to the fiducial region is
subsumed into the geometric and kinematic acceptances, calculated from
the Monte Carlo simulation of the detector, as described below).

\begin{figure}
\centerline{
\psfig{figure=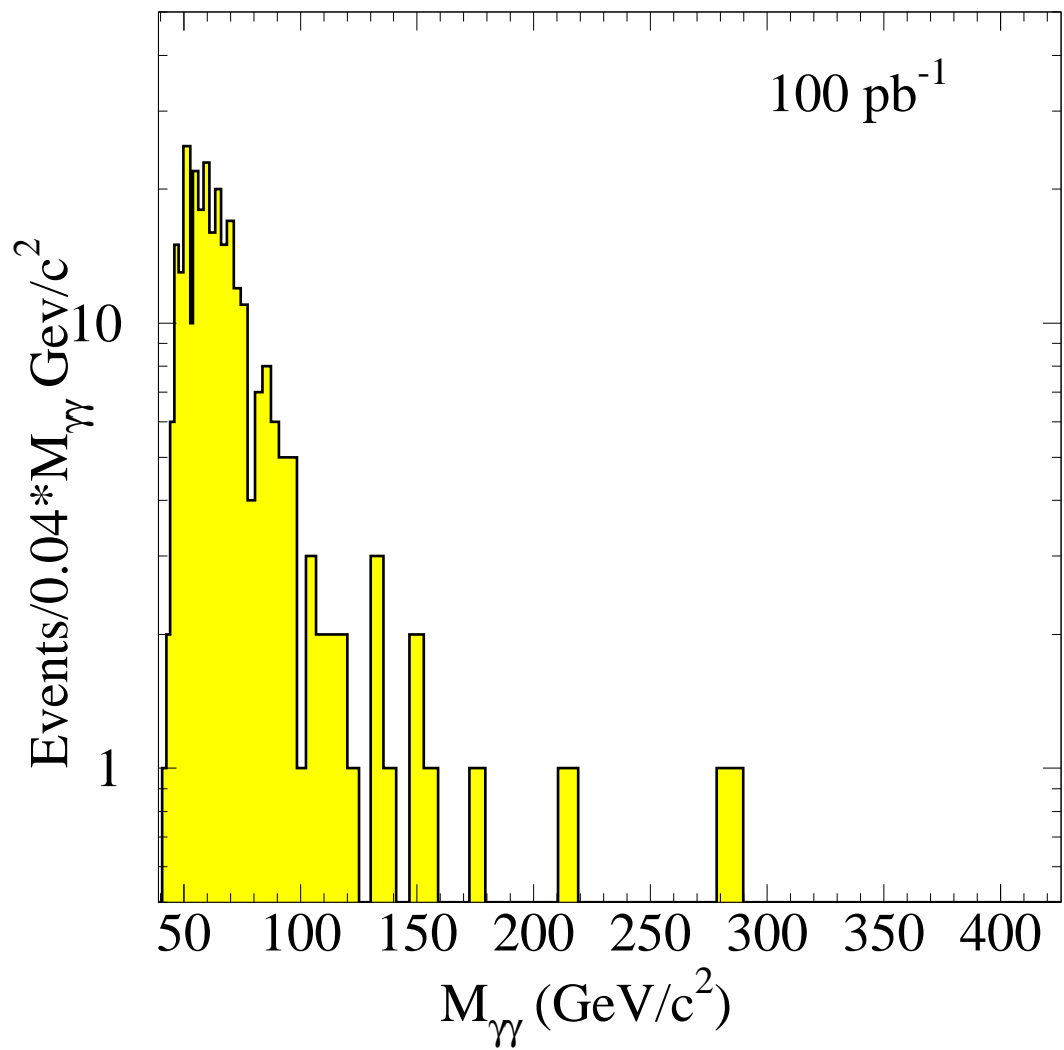,height=16cm,width=16cm}}
\caption{ The invariant mass distribution of diphoton candidates (287 events) 
with a
bin-width of 4\% of the mass. Note that the three highest-mass bins contain one
event each.}
\label{fig:mass}
\end{figure}

Figure~\ref{fig:mass} shows the invariant mass distribution of the 287
diphoton candidate events that pass the selection criteria.  A variable
bin-width has been chosen to correspond to two times the mass
resolution (2$\sigma$) to enable the observation of narrow structures.

\section{Backgrounds}

The dominant backgrounds for this analysis are $\gamma-$jet and
jet-jet production, where the jets have `faked' photons by fluctuating
to a single $\pi^0$\ or $\eta$, and real photon pairs from prompt QCD
production.  The estimated background from $Z^0\to e^+e^-$\ with both
electrons faking photons is less than 1 event.
 
The jet fake rate is measured directly from the data using methods
developed for measurements of the inclusive photon ~\cite{photonxsec}
and di-photon cross-sections ~\cite{diphoxsec}.  For clusters with
$E_T<35$~GeV, the lateral shape of the shower in the CES system is
used to discriminate between prompt photons and photons from
$\pi^0\to\gamma\gamma$.  Above 35~GeV, where the shapes of showers in
the CES from photons and $\pi^0$s are indistinguishable, the
difference in conversion probability of a single photon and a pair of
photons (from $\pi^0$ decay) in the material of the magnet coil in
front of the CPR chambers is used to calculate the single-photon
purity.  These probabilities are used to calculate weights for each
event being `photon-photon', `photon-fake' or `fake-fake'.  The result
of applying this method to the sample of 287 event diphoton candidates
is that $183\pm56\pm32$\ events are `photon-fake' or `fake-fake'. This
corresponds to a background fraction of $64\pm19\pm11\%$, where the
first uncertainty is statistical and the second is systematic.  The
systematic uncertainty comes primarily from uncertainties in the
modeling of the back-scattering of photons from the electro-magnetic
shower in the calorimeter into the CPR chambers, and the modeling of
the shower shapes in the CES chambers.

The mass spectrum of the jet fakes is determined using a control
sample of events enriched in fake photons.  This sample is made using
the same selection requirements as the diphotons except that one or
both clusters fail the calorimeter isolation requirement.  This sample
contains some real diphotons which fail the isolation requirement.  
From studies of high-$E_T$ electrons from W and Z decays, we estimate that
10\% of diphoton signal events will end up in the non-isolated sample.
  The mass
distribution of the 198 event non-isolated sample is normalized to the
number of fake events measured in the diphoton candidate sample (183
events).

Two standard model processes make significant contributions to prompt
diphoton production: $q \bar q \rightarrow \gamma \gamma$\ and $gg
\rightarrow \gamma\gamma$.  In addition, initial and final state
electromagnetic radiation from $\gamma-$jet production contributes
indirectly to the diphoton mass spectrum.  In the indirect case,
several processes contribute to $\gamma-$jet production: $q \bar q
\rightarrow g \gamma $, $qg \rightarrow q\gamma$, and $qq \rightarrow
g\gamma$.  These standard model processes are modeled using the Monte
Carlo program PYTHIA~\cite{Pythia} with CTEQ4L structure
functions~\cite{CTEQ4} and the CDF fast detector simulation.  The
$\gamma\gamma$\ event selection efficiency is determined using the MC
and detector simulation, with a correction factor of 
$C_{MC}\equiv 0.76\pm 0.08$\
applied to account for differences between the detector simulation and
the actual detector performance.  These differences are dominated by
effects from additional low energy tracks from the underlying event and
from track reconstruction.  The correction factor is obtained by
comparing the efficiency of the photon selection requirements when applied
to electrons from $Z^0\to e^+e^-$\ events from Monte Carlo and data.
The $Z^0\to e^+e^-$\ events are selected with very loose requirements
to minimize any bias in the method.  A global systematic uncertainty
of $13-16\%$\ applies to these estimates, coming from the uncertainty
on the correction factor (10\%), the modeling of QED radiation (10\%
for diphoton masses below 120 GeV/c$^2$\ and 5\% above), the
dependence on the structure functions (5\%), and the integrated
luminosity (4\%).

\par
\begin{figure}
\centerline{
\psfig{figure=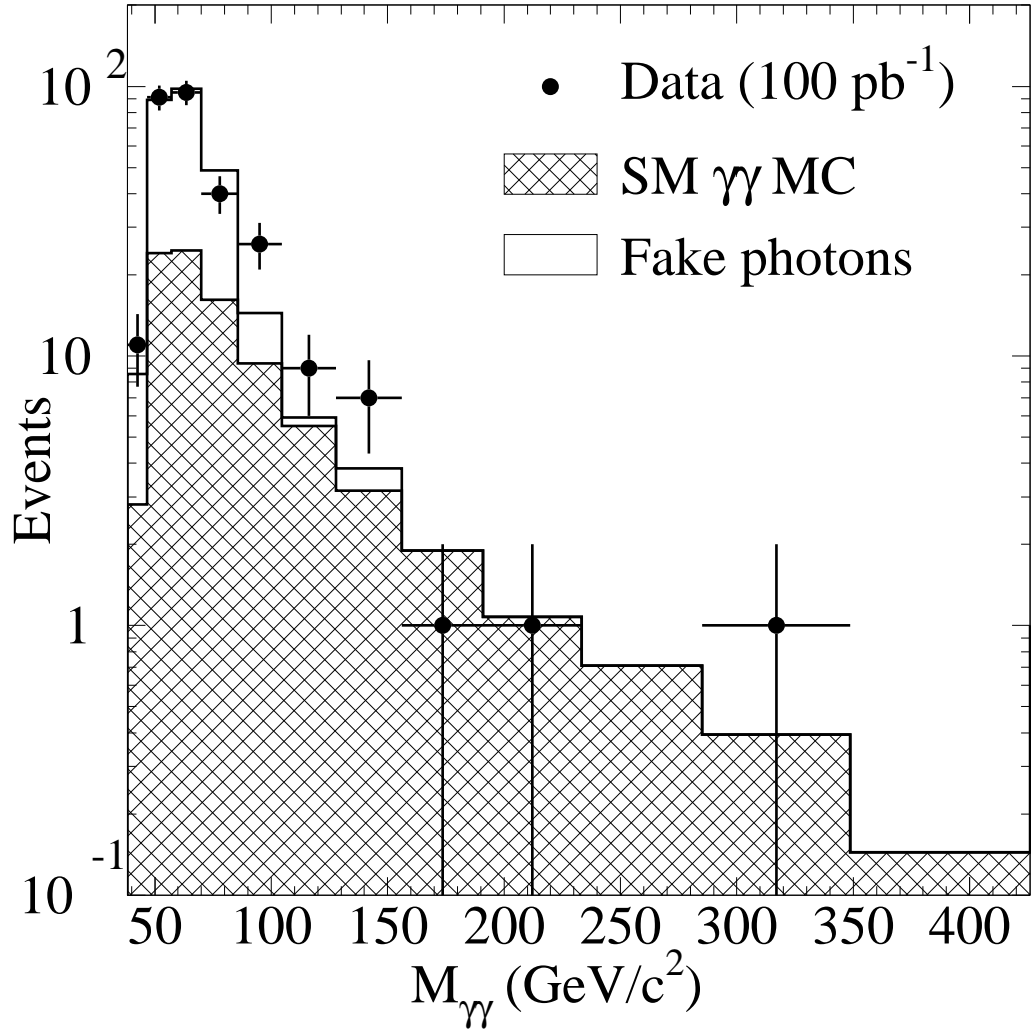,height=16cm,width=16cm}}
\caption{The diphoton
  candidate mass distribution is compared with background predictions
  with a bin-width of 20\% of the mass.  The shaded distribution
  represents the Monte Carlo prediction for QCD diphoton production;
  the unshaded distribution represents the predicted yield for jets
  faking photons.}
 \label{fig:mass2}
\end{figure}

The total predicted background from fake photons plus QCD diphoton
production is $280\pm 66$~events.  Figure~\ref{fig:mass2} shows a
comparison of the diphoton mass spectrum for the 287 isolated diphoton
candidates (points) with background predictions.  The shaded
distribution represents the standard model diphoton prediction from
the PYTHIA Monte Carlo program, while the unshaded distribution
represents the predicted spectrum from jets faking photons. The
bin-width in this plot corresponds to about 10 times the mass resolution; 
any narrow-width resonance would be seen in the finer binning of
Figure~\ref{fig:mass}.  The data are well-modeled by the background
predictions: above 70 (100) GeV/c$^2$\ we observe 85 (21) events
compared to a background prediction of $77.1\pm 15.7$\ ($14.7\pm 3.2$)
events.  The numbers of events and backgrounds are summarized in
Table~\ref{tab:inc_xsec_limit}.

\section{ Limit on Inclusive $\gamma\gamma$\ Production}

We first consider the signature of $\gamma\gamma$ +X. We set limits on
the cross section for narrow resonances with mass greater than 70
GeV/c$^2$~\cite{narrow}.  
The acceptance for diphoton production is evaluated using
the diphoton decay of a narrow resonance, $\phi\to\gamma\gamma$, as a
model of the kinematics for the production and decay of a heavy
sgoldstino.  The  sgoldstino samples are generated using the
PYTHIA Monte Carlo generator with CTEQ4M structure
functions~\cite{CTEQ4}, simulated using the CDF fast detector
simulation, and passed through the same event selection criteria as
the data. The product of \mbox{efficiency times acceptance} increases
 from 10\% at 75~ GeV/c$^2$ to 16\% at 400~
GeV/c$^2$. The correction factor $C_{MC}$ discussed above is applied
to the $\gamma\gamma$\ efficiency.  The acceptance has an additional
systematic uncertainty of 4\% due to the dependence on the structure
functions.
\par
\begin{figure}
\centerline{
\psfig{figure=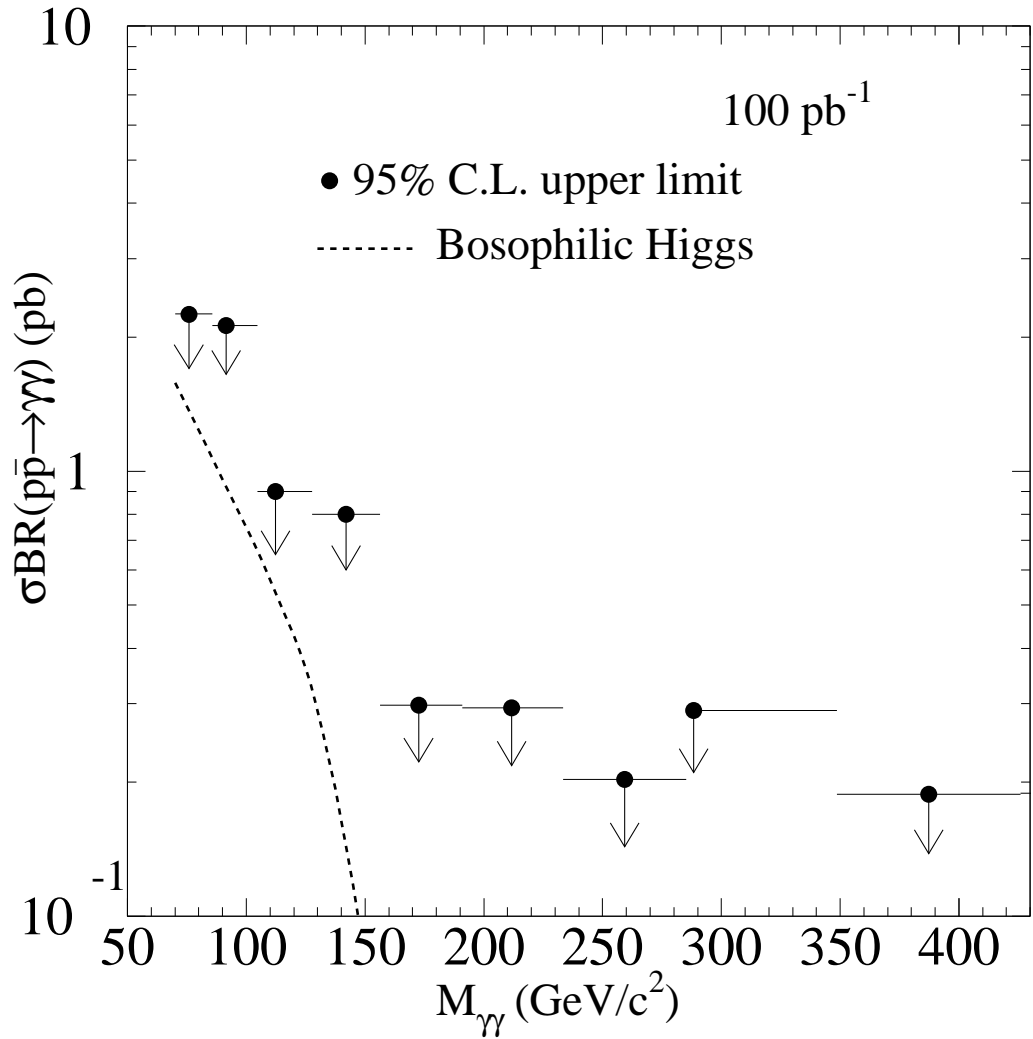,height=16cm,width=16cm}}
 \caption{Cross section limit at 95\% C.L. 
for high mass diphoton production from a resonant
state with negligible natural width. The points represent the average
mass of the events in each bin, 
but the limits are evaluated at the bin center. 
The theoretical cross section for a bosophilic Higgs boson~\protect\cite{stange94}
  is shown as a
dashed line.}
 \label{fig:xsec_limit}
\end{figure}

 The cross section limit in each mass bin of Table \ref{tab:inc_xsec_limit}
above 70 GeV/c$^2$ is given by the following expression:
\begin{equation}
   \sigma(p \bar p \rightarrow \gamma \gamma) < {{N^{95\%CL} (\gamma\gamma)}
                        \over {\epsilon \cdot \ A \cdot \int {\cal L} dt}}
   \label{eqn:xsec_limit}
\end{equation}
where $N^{95\%CL}(\gamma\gamma)$\ is the 95\% C.L.
upper limit on the number of diphoton events in the mass bin,
$\epsilon$\ is the selection efficiency, $A$\ is the acceptance
evaluated in the center of the bin, and $\int {\cal L} dt$\ is the
integrated luminosity.  The upper limit on the number of events in
each bin is determined using a Monte Carlo technique ~\cite{gluinos}
which convolutes the uncertainties (including systematic
uncertainties) on acceptance, efficiency and the integrated luminosity
with the background expectations. The total systematic uncertainty of 12\%
consists of 4\% from the luminosity measurement, 10\% from
the selection efficiencies, and 4\%  from the
acceptance. Table~\ref{tab:inc_xsec_limit} provides a summary of the
limits.  Figure~\ref{fig:xsec_limit} shows the cross section limits in
nine mass bins above 70 GeV/c$^2$.  For comparison, the cross section
times branching ratio for $p\bar p \to H^0+W/Z \to \gamma\gamma+X$\
production is shown (dashed curve) for bosophilic branching
ratios~\cite{stange94}. The curve corresponding to the standard model
branching ratio is not shown, being at least one order of magnitude
below the bosophilic one.

\subsection{Limits on the production of heavy sgoldstinos}
In the scenario in which squarks, sleptons, gluinos, charginos,
neutralinos and Higgs bosons are sufficiently heavy not to play any r\^ole
 in
sgoldstino decays, the most important decays are the two-body decays:
$\phi\to {\tilde G}{\tilde G},\gamma\gamma,gg,\gamma Z,ZZ,W^+W^-$ and 
$f\bar
f$. Three and four-body decays are also possible but quite
suppressed.  Sgoldstino couplings can be parameterized in terms of the
supersymmetry-breaking scale $\sqrt{F}$, the gaugino masses, $M_3$,
$M_2$\ and $M_1$, and a mass parameter, $\mu_a$, associated with the
charged higgsino. To account for the $t\bar t\phi$\ coupling for heavy
sgoldstinos, two arbitrary free parameters, $A_S$\ and $A_P$, with the
dimension of a mass are introduced.  We adopt in the following the two
sets of choices for the parameters adopted in Ref.~\cite{zwire}: these
choices represent a situation in which sgoldstino production is more
important than gluino/chargino/neutralino production. The two sets
correspond to chargino masses of about (220, 380) for case A and about
(270, 430) GeV/c$^2$\ for case B.
\par
In order for the calculations to be valid, the sgoldstino total width
has to be small compared to $m_\phi$.  For both parameter sets the decay
$\phi\to gg$\ dominates, but $\phi\to\gamma\gamma$\ is not negligible,
being of the order of few percent.

\par
The dominant mechanism for sgoldstino production is gluon-gluon fusion
$g+g\to\phi$, while other associated processes such as $q+\bar q\to
V+\phi$ ($V=\gamma,W,Z)$ or $q+\bar q\to q+\bar q+\phi$\ are
suppressed by about four orders of magnitude. The calculation of the
production cross section has been made at lowest order
(LO)~\cite{zwirh}; however NLO QCD corrections to $\sigma(p\bar
p\to\phi)\times BR(\phi\to\gamma\gamma)$\ are expected to be
negligible because they have cancelling effects in the cross section
and branching ratio.
\par
\begin{figure}\centerline{
\psfig{figure=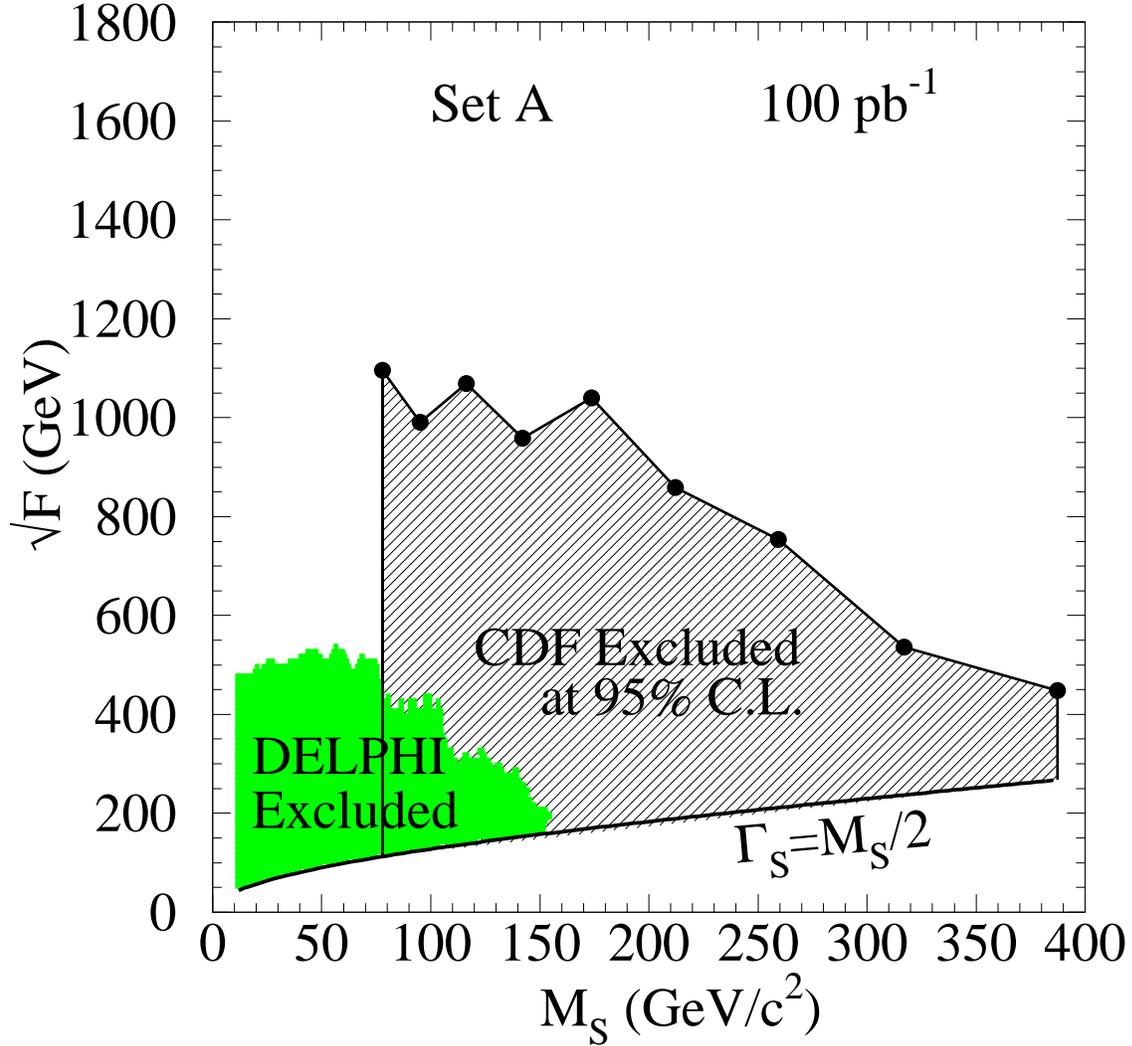,height=16cm,width=16cm}}
  \caption{The exclusion region at the 95\% C.L. for S-type (CP-even)
sgoldstinos in the 
$M_S$-$\sqrt{F}$ space for the parameters of set A: $M_3=A_S=A_P=400$, 
$M_2=\mu_a=300$, $M_1=200$ GeV/c$^2$. 
$M_S$ is mass of the S-type sgoldstino.
The CDF results are shown as the
 hatched area; the region excluded by results from DELPHI~\protect\cite{delphi}
is shown as the solid shaded area. The points represent the mass at which
the limits are calculated.
The  boundary $\Gamma_S=M_S/2$ beyond which the model may not be valid
is also shown.}
  \label{fig:flim_a}
\end{figure}
\begin{figure}
\centerline{
\psfig{figure=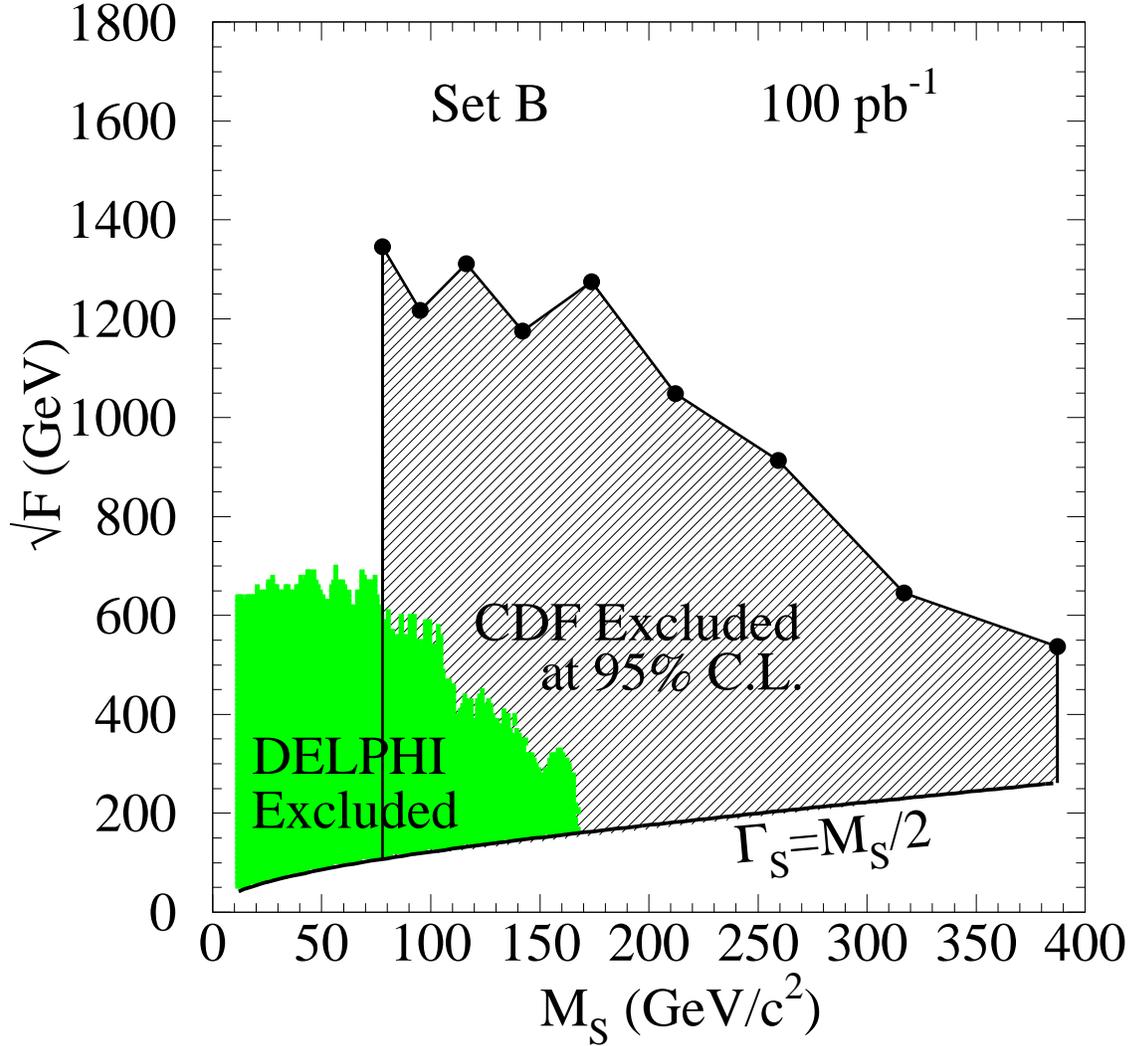,height=16cm,width=16cm}}
  \caption{Exclusion region at the 95\% C.L. in the 
$M_S$-$\sqrt{F}$\ space for the parameters of set B: 
$M_3=M_2=M_1=\mu_a=A_S=A_P=350$ GeV/c$^2$. 
The CDF results are shown as the
 hatched area; the region excluded by results from DELPHI~\protect\cite{delphi}
is shown as the solid shaded area.  The points represent the mass at which
the limits are calculated.
The  boundary $\Gamma_S=M_S/2$ beyond which the model may not be valid
is also shown.}
  \label{fig:flim_b}
\end{figure}

Comparing the limits found on the inclusive production 
cross section to the  theoretical value of
$\sigma(p\bar p\to\phi)\times BR(\phi\to\gamma\gamma)$\ bin-by-bin, 
and considering its
 $1/F^2$ dependence, we derive lower limits on $\sqrt{F}$\ for sgoldstino
masses corresponding to the center of the bin. These limits are represented
as exclusion regions in the $M_\phi$\ vs $\sqrt{F}$\ space. Figures 
\ref{fig:flim_a} and \ref{fig:flim_b} show
these limits for the $S$-type sgoldstinos. The limits  for the $P$-type
(CP-odd)  sgoldstino are very similar, differing by less than 0.1\%.
No limit is set in the region $\Gamma_\phi > M_S/2$, where the
theoretical calculation may not be valid~\cite{zwirh}. 

\section{ Selecting $\gamma\gamma$ + W/Z Candidates} 
The inclusive $\gamma\gamma$\ analysis is not sensitive to production
of a bosophilic Higgs decaying to two photons in the lower-mass region 60-100
GeV/c$^2$\ because the backgrounds from jets faking photons and QCD
diphoton production are too high (see Figure~\ref{fig:xsec_limit}).
To increase sensitivity in this mass region we narrow the signature to
be $\gamma\gamma + W/Z$.  The additional requirement of a $W$\ or $Z$\
boson significantly reduces these backgrounds, allowing access to
smaller cross sections.
 
To achieve a high acceptance for all $W$\ and $Z$\ decay channels, the
vector bosons are selected using simple signatures which yield
significant background reductions without the inefficiency of full
reconstruction.  The backgrounds from jet fakes and QCD
$\gamma\gamma$\ production are evaluated using the non-isolated sample
of 198 events and PYTHIA Monte Carlo QCD background sample used in the
inclusive $\gamma\gamma$ analysis previously described.  Backgrounds
from electroweak processes are found to be insignificant.
\par

The vector boson selection consists of the logical OR of three general 
categories based on decay channels as follows:
\begin{enumerate}
  \item Central isolated electron ($E_T >20$~GeV) or muon ($P_T > 20$~ GeV/c) 
        for $W\to l\nu$\  and $Z^0\to l^+l^-$
  \item Two Jets ($E_T>15$~GeV, $|\eta|<2.0$) for $W\to qq^{\prime}$\  
and $Z^0 \to q \bar q$
  \item $\ebar_T> 20$~GeV for $W\to l \nu$\  and $Z^0\to\nu\bar \nu$
\end{enumerate}
where $\ebar_T$\ is the symbol for missing transverse energy~\cite{met}.

Leptons ($e$ and $\mu$) are selected using the isolated central lepton
requirements used in the `lepton-plus-jets' analysis for the discovery
of the top quark~\cite{top}.  The lepton identification efficiencies
are measured in data samples of $Z$\ bosons decaying to electrons
($77.8\pm0.6\%$) and muons ($90.6\pm0.5\%$).    Jets are identified in the
calorimeter using a fixed cone algorithm~\cite{jetalg} with a cone size
in $\eta$-$\phi$\ space of radius $\Delta R=0.4$.  Any jet within a
radius of 0.4 in $\eta$-$\phi$\ space of an electron or within a radius
of 0.6 of a photon is ignored.  Finally the jet-jet invariant mass is
required to be consistent with a $W$\ or $Z$\ boson: $40<M_{JJ}<130$~
GeV/c$^2$.
The missing transverse
energy is corrected for any high-$P_T$\ central muons.  Since
mismeasured jet energies can result in false $\ebar_T$, events with
any jet ($E_T^{\rm jet}>10$~GeV) within $25^{\circ}$\ of the
$\ebar_T$\ direction are rejected. The same exclusion applies for events
with $\ebar_T$ near photons ($E_T^\gamma > 22$ GeV), electrons 
($E_T^e >20$ GeV) and muons ($P_T^\mu> 20$ GeV/c).

The results of the $\gamma\gamma + W/Z$\ event selection are
summarized in Table~\ref{tab:event_summary} listing the number of
events satisfying each $W/Z$\ selection requirement.  Some properties
of the 6 events passing the selection requirements are listed in
Table~\ref{tab:vh_events} including one event which passes both the
jet-jet and $\ebar_T$\ selection requirements.  The highest mass event
has a $\gamma\gamma$\ invariant mass of 137~GeV/c$^2$\ and
$\ebar_T=21$~GeV.  The total estimated background for
$M_{\gamma\gamma}>130$~GeV/c$^2$\ is $0.19\pm0.12$\ events, due to
standard model $\gamma\gamma$ production.

\begin{figure}
\centerline{
\psfig{figure=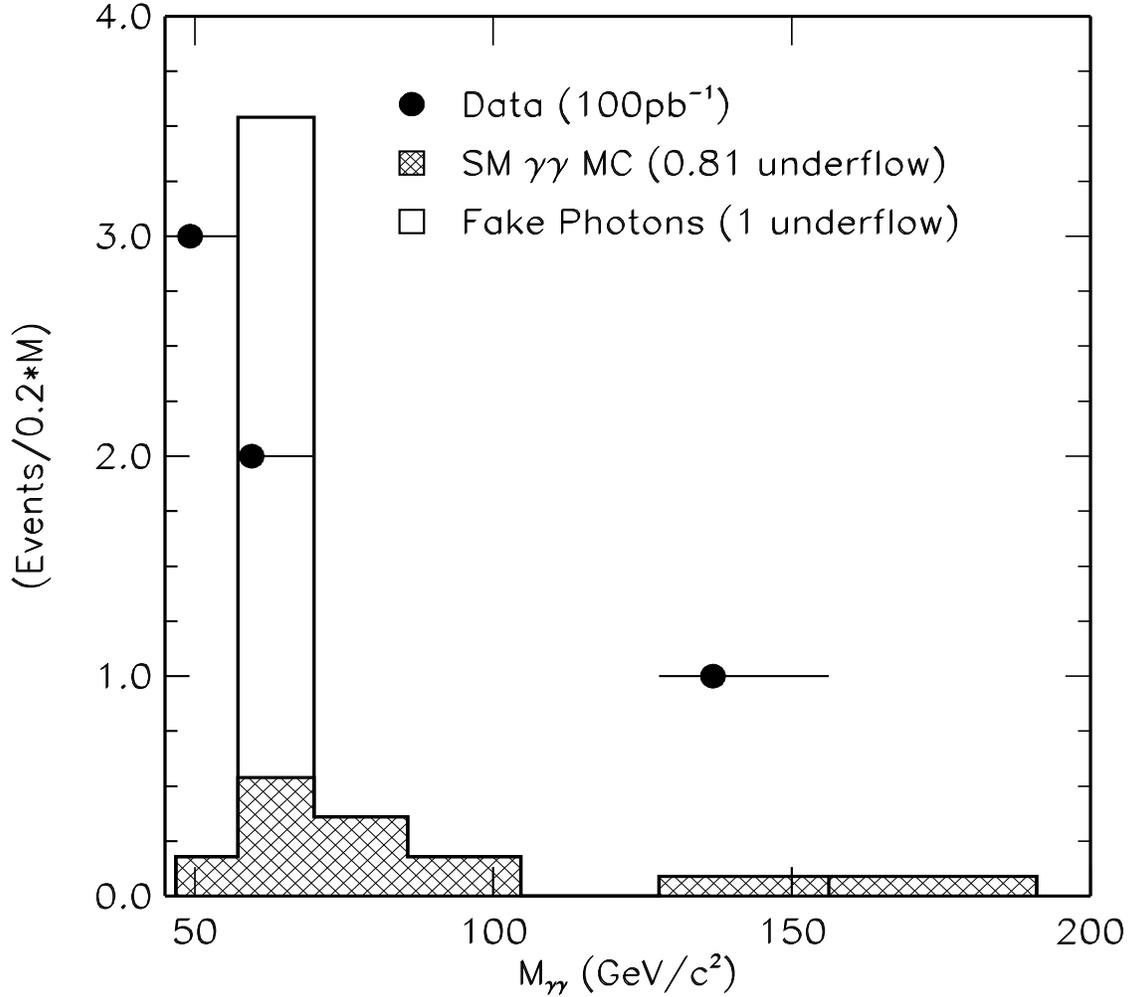,height=16cm,width=16cm}}
  \caption{Photon-photon mass distribution compared with background predictions
 	for events passing the $\gamma\gamma + W/Z$\  selection. The 
        cross-hatched
	distribution represents the Monte Carlo prediction for QCD diphoton 
	production; the shaded one represents the predicted yield from jets 
	faking photons.  The choppy appearance of the background estimates
	is the result of low efficiency for the $W/Z$\ selection.  The small
	electroweak backgrounds are not shown.}   
  \label{fig:data_mgg}
\end{figure}

Table~\ref{tab:event_summary} also lists the estimated backgrounds
from photon fakes, QCD $\gamma\gamma$\ production, and electroweak
sources, which total $6.4\pm2.1$ events.  Fake-photon backgrounds,
which are estimated from the non-isolated data sample, contribute 1
event to the $\ebar_T$\ category and 3 events to the jet-jet category.
Backgrounds from QCD $\gamma\gamma$, which are estimated using a
sample generated with the PYTHIA Monte Carlo equivalent to 1~fb$^{-1}$
of data, contribute $0.8$\ events to the $\ebar_T$\ category and
$1.6$\ events to the jet-jet category\footnote{There is a small overlap
between signatures for the QCD $\gamma\gamma$\ background.}.  There are
small electroweak backgrounds, $0.2\pm0.2$\  events which contribute to the
electron signature from events with a $W$\ or $Z$\ boson produced in
association with multiple photons and/or jets.  These events only
contribute in the case where the $W(Z)$\ decays to an electron(s) and
the charged track(s) associated with the electron(s) is not
reconstructed.  Figure~\ref{fig:data_mgg} shows the $\gamma\gamma$\
mass distribution of events passing all $\gamma\gamma + W/Z$\
selection for the isolated diphoton data and the background samples.
The mass distribution for the electroweak events is neglected.  There
is no evidence of a $\gamma\gamma$\ resonance in the data.

\section{ Limits on $\gamma\gamma$ + W/Z Production}

We set an upper limit on the cross section times branching ratio for
the process $p \bar p \to  \gamma\gamma + W/Z$  as a function of $\gamma\gamma$
mass:
\begin{equation}
   \sigma(p \bar p \rightarrow \gamma \gamma + W/Z) < 
                        {{N^{95\%CL} (\gamma\gamma + W/Z)}
                \over   {\epsilon \cdot \  A \cdot \int {\cal L} dt}}
   \label{eqn:vh_xsec_limit}
\end{equation}
where $N^{95\%CL} (\gamma\gamma + W/Z)$\ is the 95\% C.L. 
upper limit on the
number of events, $\epsilon \cdot A$\ is the product of efficiency
times acceptance, and $\int {\cal L} dt$\ is the integrated
luminosity.  The number of signal events at each mass is taken as the
number of isolated diphoton events passing the vector ($W/Z$)
selection cuts and falling within a $\pm 3\sigma(M_H)$\ mass window
around the candidate mass, $\sigma$ being about 2 (3) GeV/c$^2$ for 
$M_H=100$ (150) GeV/c$^2$.  We calculate $N^{95\%CL}$ at each mass, 
assuming no
background subtraction and including a Gaussian systematic uncertainty
of 15\% which includes diphoton selection efficiency (10\%), luminosity (4\%),
gluon radiation modeling (11\%), and jet energy scale (7\%).

\begin{figure}\centerline{
\psfig{figure=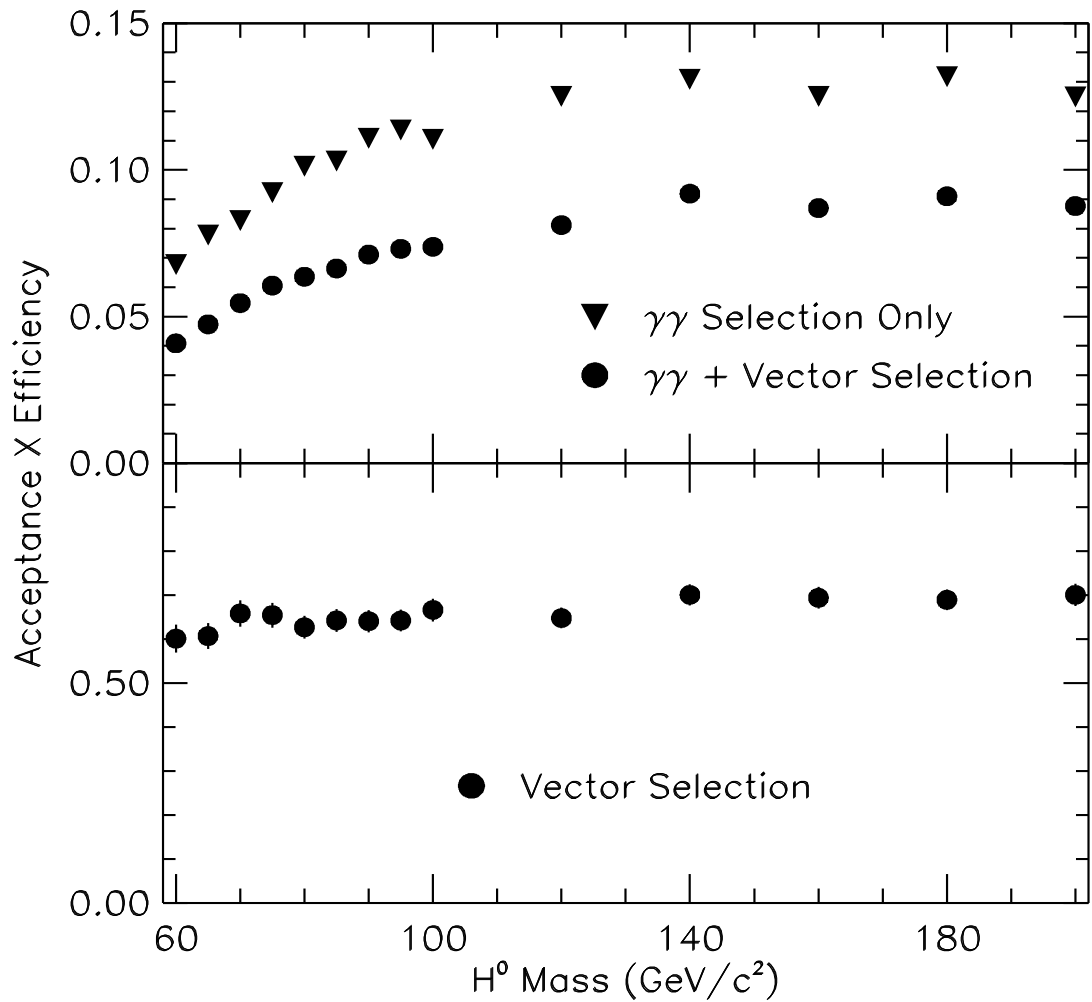,height=16cm,width=16cm}}
    \caption{Acceptance$\times$efficiency for $VH$\ production, with the $W$\ 
and $Z$\ bosons decaying via any SM decay and the Higgs boson 
decaying to $\gamma\gamma$.}
    \label{fig:vh_acceptance}
\end{figure}

The acceptance is determined from Monte Carlo samples of associated 
Higgs + W/Z generated with PYTHIA and CTEQ4L structure functions~\cite{CTEQ4}.
Figure~\ref{fig:vh_acceptance} shows the product of the efficiency
times acceptance as a function of $M_H$\ before and after the vector
boson selection cuts.
The efficiency times acceptance increases from  about 4\%
 for $M_H=60$~ GeV/c$^2$ 
to  about 9\% for $M_H>100$~ GeV/c$^2$. 
 The mass dependence of the acceptance is 
dominated by the photon $E_T$ requirement.
\par
\begin{figure}
\centerline{
\psfig{figure=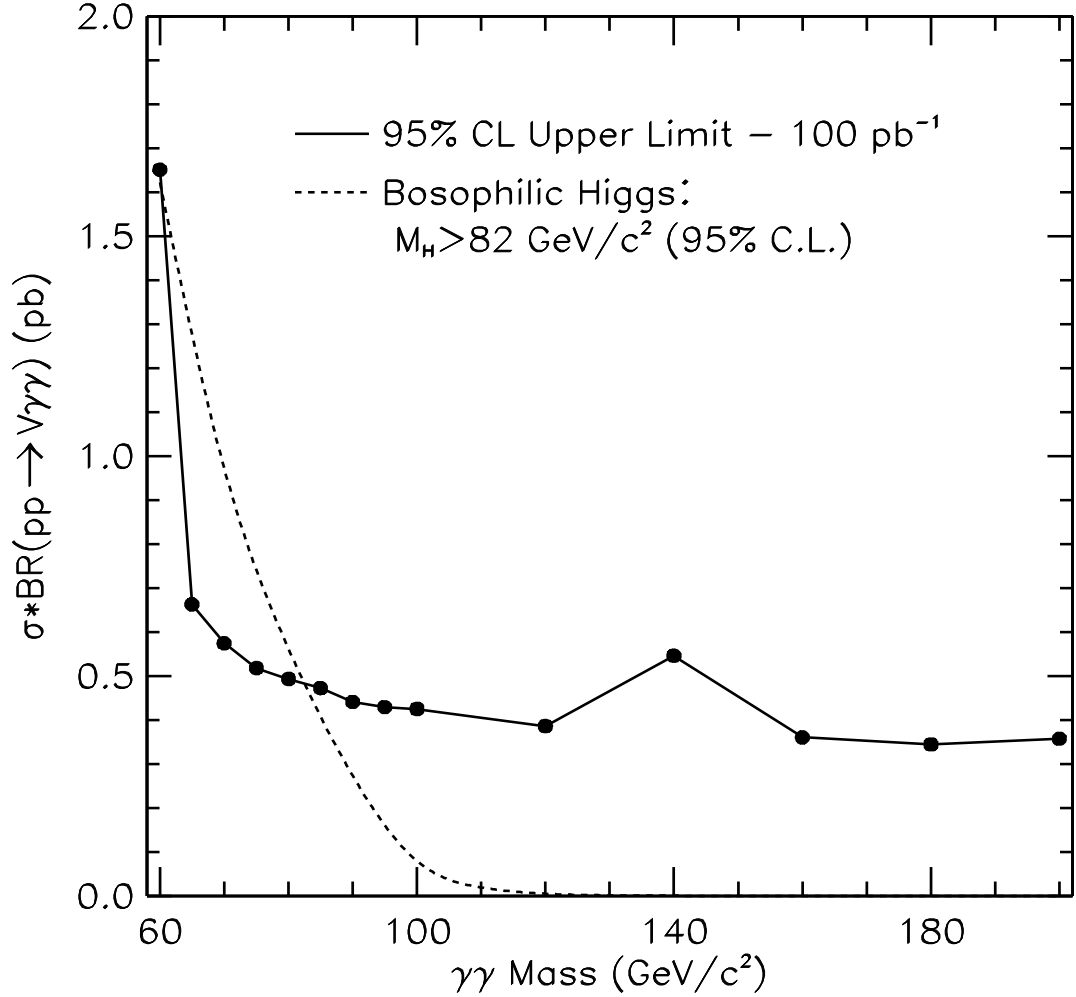,height=16cm,width=16cm}}
  \caption{Upper limit at 95\% C.L. on the $\gamma\gamma + W/Z$\  cross 
	section as a function of $\gamma\gamma$\ mass.  The dashed curve 
	shows the prediction for cross section times branching 
	fraction for a bosophilic $H\to \gamma\gamma$\ with branching fraction
	from reference~\protect\cite{stange94} and the cross section for associated 
	Higgs production is a Standard Model NLO calculation from 
	reference~\protect\cite{higgs_nlo_xsec}.}
  \label{fig:vh_sbr_limit}
\end{figure}

Figure~\ref{fig:vh_sbr_limit} shows the 95\% C.L. upper limit on the
cross section times branching ratio for $p \bar p \to \gamma\gamma + W/Z$.
The overlayed dashed curve is the prediction for a bosophilic Higgs
using the branching ratios from reference~\cite{stange94} and a NLO
cross section calculation from reference~\cite{higgs_nlo_xsec}, using
the CTEQ4M structure functions~\cite{CTEQ4}.  
A 95\% C.L. lower limit on the mass
of a bosophilic Higgs is set at 82 GeV/c$^2$.
Table~\ref{tab:wh_xsec_limit} provides a
summary of the limit.
\par
\begin{figure}
\centerline{ 
\psfig{figure=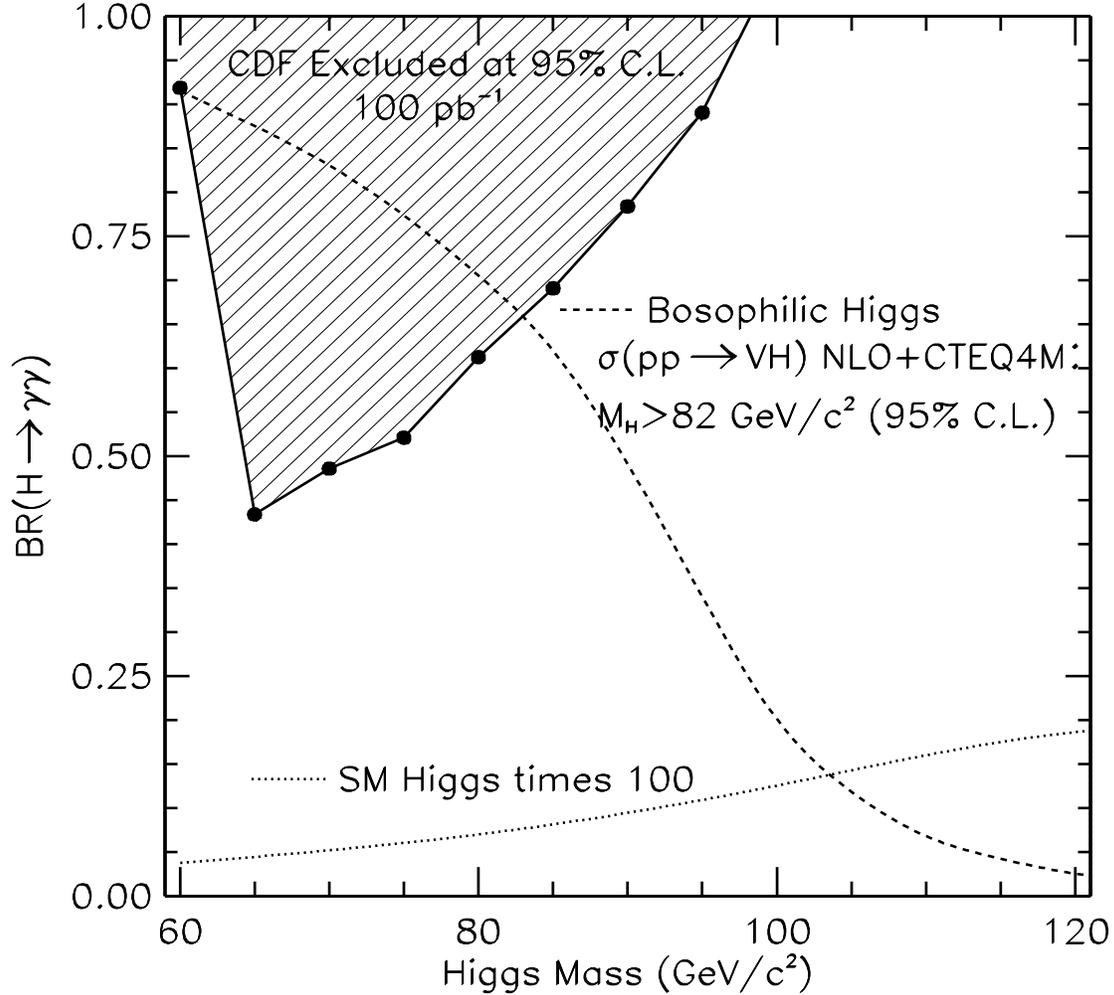,height=16cm,width=16cm}}
  \caption{Upper limit at 95\% C.L. on the branching ratio for 
	$H\to \gamma\gamma$\ assuming standard model production for 
	$W/Z + H$~\protect\cite{higgs_nlo_xsec}. 
 Note that the limit lies within the regions excluded by
 OPAL~\protect\cite{opal_ggx} and ALEPH~\protect\cite{aleph_ggx}.
        The dashed curve 
	shows the branching fraction for a bosophilic $H\to \gamma\gamma$
	from reference~\protect\cite{stange94}.}
  \label{fig:vh_br_limit}
\end{figure}

An upper limit on the branching fraction for $H\to\gamma\gamma$  is obtained
by dividing the cross section limit on $\gamma\gamma + W/Z$  by the predicted 
cross section for $W/Z+H$  production.
The resulting branching ratio upper limit is shown in
Figure~\ref{fig:vh_br_limit}, and lies within the regions excluded by 
OPAL~\cite{opal_ggx} and ALEPH~\cite{aleph_ggx}.
The overlayed dashed and dotted curves are the 
predictions for a bosophilic and Standard Model Higgs boson
(scaled up by a factor of 100), respectively.

\section{ Conclusions}

We have presented results of searches for massive diphoton production both
inclusively and in association with a high-$P_T$ lepton, $\ebar_T$, or
dijets.  The latter channels are sensitive to production of a vector
boson in association with a Higgs boson which subsequently decays to
photons.  Both the inclusive and exclusive signatures are consistent
with predictions from standard model sources.
In the inclusive channel we set upper limits on the production of 
narrow resonances decaying into two photons. Comparing these limits to a
LO calculation for massive sgoldstino production we set limits in the
range of 1 TeV on the
supersymmetry-breaking scale $\sqrt{F}$ for two sets of parameters.
In the exclusive channels,  we set an upper limit on the cross
section times branching fraction for $p \bar p \to \gamma\gamma + W/Z$
between 60 and 200~ GeV/c$^2$.    Using a NLO
calculation of the SM cross section for $p\bar p \to VH$ we set a
95\% C.L. upper limit on the branching ratio for $H\to \gamma\gamma$.
Between approximately 60 and 100~ GeV/c$^2$ the upper limit on the
branching ratio is less than 1.  Using the branching ratios of 
reference~\cite{stange94} the lower limit on the mass of a
bosophilic Higgs is 82 GeV/c$^2$ at 95\% C.L.

\section{ Acknowledgments}

     We thank the Fermilab staff and the technical staffs of the
participating institutions for their vital contributions.  This work was
supported by the U.S. Department of Energy and National Science Foundation;
the Italian Istituto Nazionale di Fisica Nucleare; the Ministry of Education,
Science, Sports and Culture of Japan; the Natural Sciences and Engineering
Research Council of Canada; the National Science Council of the Republic of
China; the Swiss National Science Foundation; the A. P. Sloan Foundation; the
Bundesministerium fuer Bildung und Forschung, Germany; the Korea Science
and Engineering Foundation (KoSEF); the Korea Research Foundation; and the
Comision Interministerial de Ciencia y Tecnologia, Spain.
\clearpage


\newpage
\raggedright

\newpage

\begin{table}
 \begin{center}
  \begin{tabular}{|c|c||c|c|c|c|c|} 
Mass  & Events & Fake    & SM         & Total & $\epsilon A$ & $\sigma$\\
(GeV/c$^2$)    & Photons & Production &  &  & & (pb) \\
\hline
46.8-57.2 & 90 & $65.2\pm 23.8$ & $24.1\pm 3.9$ & $89.3\pm 24.1$ & 0.04 & --\\
\hline
57.2-70.0 & 95 & $73.3\pm 26.7$ & $24.6\pm 3.9$ & $97.9\pm 27.0$ &  0.07 & --\\
\hline\hline
70.0-85.6  & 40& $32.6\pm 12.5$ & $16.2\pm 2.6$ & $48.8\pm 12.7$ & 0.107 & 2.25\\
\hline
85.6-104.6  & 26 & $5.0\pm 2.6$ & $9.4\pm 1.5$ & $14.4\pm 3.0$ & 0.112 &2.12\\
\hline
104.6-127.8  & 9  & $0.4^{+1.0}_{-0.4}$ & $5.5\pm 0.9$ & $5.9^{+1.3}_{-1.0}$ 
& 0.119 & 0.90\\
\hline
127.8-156.2  & 7 & $0.6^{+1.0}_{-0.6}$ & $3.2\pm 0.4$ & $3.8^{+1.1}_{-0.7}$ 
& 0.126 & 0.80\\
\hline
156.2-191.0  & 1 & $<0.04$ & $1.9\pm 0.3$ & $1.9\pm 0.3$ & 0.134 & 0.30\\
\hline
191.0-233.4  & 1 & -- & $1.1\pm 0.2$ & $1.1\pm 0.2$ & 0.143 & 0.29\\
\hline
233.4-285.2  & 0  & -- & $0.7\pm 0.1$ & $0.7\pm 0.1$ & 0.151 &0.20\\
\hline
285.2-348.6  & 1  & -- & $0.4\pm 0.1$ & $0.4\pm 0.1$ & 0.158 & 0.29\\
\hline
348.6-426.0  & 0 & -- & $0.1\pm 0.1$ & $0.1\pm 0.1$ & 0.163 & 0.19\\
\hline\hline
Total  & 270& $177.1\pm 62.3$ & $87.2\pm 14.4$ & $264.3\pm 63.9$ &-- &-- \\
  \end{tabular}
  \end{center}
  \caption{The number of diphoton events observed, background from jets
faking photons, `background'
from standard model diphoton production, 
total background, efficiency times acceptance, and 95\% C.L. cross section 
limit for $\gamma\gamma+X$ production 
for each mass bin. Mass bins have a width of 20\% of the bin center.
The first two bins are not used for cross section limits, due to their low
acceptance.}
\label{tab:inc_xsec_limit} 
\end{table}

\begin{table}
\begin{center}
\begin{tabular}{|c|c|c|c|c|c|c|c|c|}
Set & $M_3$ & $M_2$ & $M_1$ & $\mu_a$ & $A_S$ & $A_P$ & units\\ 
\hline
A & 400 & 300 & 200 & 300 & 400 & 400 & GeV/c$^2$ \\
\hline
B & 350 & 350 & 350 & 350 & 350 & 350 & GeV/c$^2$ \\
\end{tabular}
\caption{The two sets of mass parameters used in the sgoldstino
theoretical cross section calculations.}
\label{sets}
\end{center}
\end{table} 

\begin{table}
 \begin{center}
  \begin{tabular}{|l|c|c|} 
 Selection                      & Isolated      & Background               \\
                                & Sample        & Estimate  \\
  \hline
 Two Isolated Photons $E_T^\gamma>22$~GeV
                                & 287           & $280\pm 66$            \\
\hline
 Central Electron, $E_T>20$~GeV &  1            & $0.2\pm0.2$           \\
 Central Muon, $P_T>20$~GeV/c     &  0          & 0                     \\
 $\ebar_T > 20$ GeV               &  3          & $1.8\pm1.3$           \\

 2 Jets ($E_T>15$ GeV, $40<M_{JJ}<130$ GeV/c$^2$)
                                &  3            & $4.6\pm1.9$           \\ 
                                \hline
 Any $W/Z$  signature           &  6            & $6.4\pm2.1$           \\
  \end{tabular}
 \end{center}
 \caption{Summary of the $\gamma\gamma$ + W/Z candidate events.  The number of
        $\gamma\gamma$\ candidate events passing each of the W/Z selection 
        requirements are listed.  There is one  event which passes both the 
        jet-jet and $\ebar_T$\ selection requirements.   The background 
        estimates come primarily from fakes (non-isolated control sample)
        plus SM $\gamma\gamma$\ production with a small contribution from 
        electroweak sources.  Some background events pass more than 
        one of the W/Z selections. }  
 \label{tab:event_summary}
\end{table}

\begin{table}
 \begin{center}
  \begin{tabular}{|c|c|l|c|l|} 
  Run  & Event		& Channel(s)	
				& $M_{\gamma\gamma}$ 
					& Properties		\\
  	& 		& 	& (GeV/c$^2$) 	& 		\\\hline
 45219	& 277283	& $\ebar_T$,jet-jet	
				& 59.1	& $\ebar_T=28.8$~GeV, 
					  $M_{JJ}=96.1$~GeV/c$^2$ 	\\
 60597	& 119813	& $\ebar_T$	& 136.8	& $\ebar_T=20.8$~GeV     \\
 61514  & 9698		& Jet-jet
				& 48.9	& $M_{JJ}=75.1$~GeV/c$^2$ 	\\
 68739	& 257646	& Electron	
				& 47.1	& $E_T = 36.1$~GeV \\
 68847  & 264160	& Jet-jet
				& 59.9	& $M_{JJ}=74.6$~GeV/c$^2$ 	\\
 70019	& 155639	& $\ebar_T$	& 51.7	& $\ebar_T=22.0$~GeV 	\\
  \end{tabular}
 \end{center}
 \caption{Features of the six events passing the $\gamma\gamma + W/Z$
	selection requirements. The event in the electron channel is the
$ee\gamma\gamma\ebar_T$ event~\protect\cite{ggmet}.}   
 \label{tab:vh_events}
\end{table}

\begin{table}
 \begin{center}
  \begin{tabular}{|c|c|c|} 
$M_{\gamma\gamma}$ & $\epsilon \times A$  & 
$d\sigma$/$dM_{\gamma\gamma}$ \\
$(GeV/c^2)$ &                & (pb/GeV$^2$) \\
  \hline
~60 &    0.048$\pm$0.002 &   1.65  \\
~65 &    0.047$\pm$0.002 &   0.66  \\
~70 &    0.055$\pm$0.002 &   0.57  \\
~75 &    0.061$\pm$0.002 &   0.52  \\
~80 &    0.064$\pm$0.002 &   0.49  \\
~85 &    0.066$\pm$0.002 &   0.47  \\
~90 &    0.071$\pm$0.002 &   0.44  \\
~95 &    0.073$\pm$0.002 &   0.43  \\
100 &    0.074$\pm$0.002 &   0.42  \\
120 &    0.081$\pm$0.002 &   0.39  \\
140 &    0.092$\pm$0.002 &   0.54  \\
160 &    0.087$\pm$0.002 &   0.36  \\
180 &    0.091$\pm$0.002 &   0.36  \\
200 &    0.088$\pm$0.002 &   0.36  \\
  \end{tabular}
  \end{center}
  \caption{ Diphoton mass, efficiency ($\epsilon$) times acceptance ($A$), 
	and cross section
  limit (95\% C.L.) for associated $W/Z$ + high mass diphoton production.}
  \label{tab:wh_xsec_limit} 
\end{table}


\begin{thebibliography}{99}
\bibitem{coordinates} The $z$ (longitudinal) axis is along the proton beam 
axis; $r$ is the transverse coordinate.  Pseudorapidity ($\eta$) is 
$\eta\equiv-\ln(\tan(\theta/2))$, where $\theta$ is the polar angle.  
Transverse energy is defined as $E_T = E\sin\theta$.
\bibitem{cluster} The cluster for a photon in the central electromagnetic
calorimeter consists of one calorimeter tower of $0.1$ in 
$\eta$ times $15^{\circ}$
in $\phi$ and the two calorimeter towers 
on either side in $\eta$, unless the tower
is at one of the boundaries at $\eta=0$ or  $\eta=1$, in which case
the cluster is only two towers.
\bibitem{narrow} We consider ``narrow'' to mean widths comparable to or 
less than the mass resolution.
\bibitem{met} $\ebar_T$ is defined to be the magnitude of the vector sum of
transverse energy in all calorimeter towers with $\vert\eta\vert <3.6$.
\end{thebibliography}
\end{document}